\documentstyle[12pt]{article}

\topmargin
-2cm
\textwidth
15cm
\textheight
25cm
\oddsidemargin
1cm
\evensidemargin
1cm
\begin{document}
\newcommand{\newc}{\newcommand}
\newc{\sm}{Standard
Model}
\date{August 1995}
\title{ Some Issues in Soft SUSY-breaking Terms from
Dilaton/Moduli sectors\thanks{Research supported in part by:
the US DOE, under contract DE-AC03-76SF00098 (AB); the CICYT, under
contract AEN93-0673 (LEI,CM,CS); the European Union,
under contracts CHRX-CT93-0132 and
SC1-CT92-0792 (LEI,CM,CS); the INFN under postdoctoral fellowship (AB);
the Ministerio de Educaci\'on y Ciencia, under research grant (CS).}
}

\author{A. Brignole$^1$,
 L.E. Ib\'a\~nez$^2$, C. Mu\~noz$^2$ and C. Scheich$^2$
 \\ \\
$^1$Lawrence Berkeley Laboratory, \\
Berkeley CA 94720, USA
\\  \\
$^2$Departamento de F\'{\i}sica Te\'orica C-XI, \\
Universidad Aut\'onoma de Madrid, \\
Cantoblanco, 28049 Madrid, Spain
\\  }
\maketitle

\vspace{-4.5in}

\rightline{FTUAM 95/26}
\rightline{LBL-37564}
\rightline{hep-ph/9508258}

\vspace{4.5in}

\begin{abstract}

We study the structure of the soft SUSY-breaking terms obtained
from some classes of 4-D strings
under the assumption of dilaton/moduli dominance in the process of
SUSY-breaking. We generalize previous analyses
in several ways and in particular consider the new
features appearing when several moduli fields contribute to SUSY breaking
(instead of an overall modulus $T$). Some qualitative
features indeed change in
the multimoduli case.
A general discussion for symmetric Abelian orbifolds
as well as explicit examples are  given.
Certain general sum-rules involving soft terms  of different
particles are shown to apply to large classes of models.
Unlike in the overall modulus $T$ case, gauginos may be lighter
than scalars even at
the tree-level. However,
if one insists in getting that pattern of soft terms,
these sum rules force
some of the scalars  to get negative mass$^2$.
These tachyonic masses could be a problem
for standard model 4-D strings but an advantage in the case
of string-GUTs.
We also discuss the possible effects of off-diagonal
metrics for the matter fields which may give rise
to flavour-changing
neutral currents. Different sources for the
bilinear $B$ soft term
are studied. It is found that the Giudice-Masiero mechanism
for generating a ``$\mu $-term'',
as  naturally implemented in orbifolds,
leads to the prediction $|tg\beta |=1$ at the string scale,
independently of the Goldstino direction.

\end{abstract}
\maketitle

\newpage


\section{Introduction}

Recently there has been some activity in trying to obtain information
about the structure
of soft Supersymmetry (SUSY)-breaking
terms in effective $N=1$ theories coming from
four-dimensional strings. The basic idea is to identify some $N=1$ chiral
fields
whose
auxiliary components could break SUSY by acquiring a vacuum expectation
value (vev).
No special assumption is made about the possible origin of SUSY-breaking.
Natural
candidates in four-dimensional strings are 1) the complex dilaton field
$S={{4\pi}\over {g^2}}
+ia$ which is present in any four-dimensional string and 2) the moduli fields
$T^i, U^i$ which parametrize the size and shape of the compactified variety
in models obtained by compactification of a ten-dimensional heterotic string.
It is not totally unreasonable to think that some of these fields may play an
important role
in SUSY-breaking. To start with, if string models are to make any sense, these
fields
should be strongly affected by non-perturbative phenomena. They are massless in
perturbation
theory and non-perturbative effects should give them a mass to avoid deviations
from the equivalence principle and other phenomenological problems.
Secondly, these fields are generically present in large classes of
four-dimensional
models (the dilaton in all of them). Finally, the couplings of these fields to
charged matter
are suppressed by powers of the Planck mass, which makes them natural
candidates
to
constitute the SUSY-breaking ``hidden sector'' which is assumed to be present
in phenomenological models of low-energy SUSY.

The important point in this assumption of locating the seed of
SUSY-breaking
in the dilaton/moduli sectors, is that it leads to some interesting
relationships among different soft terms which could perhaps be experimentally
tested.
In ref.\cite{BIM} three of the authors presented a systematic discussion of
the structure of soft terms which may be obtained under the assumption of
dilaton/moduli dominated SUSY breaking in some classes of four-dimensional
strings,
with particular emphasis on the case of Abelian $(0,2)$ orbifold models
\cite{orbifolds}.
We mostly considered a situation in which only the dilaton $S$ and
an ``overall modulus $T$'' field contribute to SUSY-breaking.
In fact, actual four-dimensional strings like orbifolds contain several $T_i$
moduli.
Generic $(0,2)$ orbifold models contain three $T_i$ moduli
fields (only $Z_3$ has 9 and $Z_4$, $Z_6'$ have 5) and  a maximum of three
(``complex structure'') $U_i$ fields. The use of an overall modulus $T$ is
equivalent to the assumption that the three $T_i$ fields of generic orbifold
models contribute exactly the same to SUSY-breaking. In the absence
of further dynamical information it is reasonable to expect
similar contributions from the three moduli although not necessarily exactly
the same. In any case it is natural to ask what changes if one
relaxes the overall modulus hypothesis and works with the multimoduli case.
This is one of the purposes of the present paper.

In section 2 we present an analysis of the effects of relaxing
the overall modulus assumption on the results obtained for soft terms.
In the multimoduli case several parameters are needed to specify the
Goldstino direction in the dilaton/moduli space, in contrast with
the overall modulus case where the relevant information is contained
in just one angular parameter $\theta$. The presence of more free
parameters leads to some loss of predictivity for the soft terms.
However, we show that in some cases there are certain
sum-rules among soft terms which hold independently
of the Goldstino direction.
The presence of these sum rules  cause that,
{\it on average} the
{\it qualitative} results in ref.\cite{BIM}
still apply. Specifically, if one insists e.g. in
obtaining scalar masses  heavier than gauginos
(something not possible at the tree-level
in the approach of ref.\cite{BIM}) ,
this is possible in the multimoduli case, but
the sum-rules often force some of the scalars
to get negative mass$^2$.
If we want to avoid this, we have to
stick to gaugino masses bigger than
(or of order) the scalar masses.
This would lead us back to the qualitative results
obtained in ref.\cite{BIM}.
In the case of
standard model 4-D strings this
tachyonic behaviour may be particularly problematic,
since charge and/or colour could be broken.
In the case of GUTs constructed from strings,
it may just be the signal
of  GUT symmetry breaking.
We exemplify the different type of soft terms
which may be obtained in the multimoduli case in some particular
examples, including an $SO(10)$ String-GUT.


Section 3 addresses another simplifying assumption in ref.\cite{BIM}.
There only the case of diagonal kinetic terms for the charged fields
was considered. Indeed this is the generic case in most orbifolds, where
typically some discrete symmetries (or $R$-symmetries) forbid
off-diagonal metrics for the matter fields. On the other hand there are
some orbifolds in which off-diagonal metrics indeed appear and
one expects that in other compactification schemes such metrics may also
appear. This question is not totally academic since, in the presence of
off-diagonal metrics, the soft terms obtained upon SUSY-breaking are
also in general off-diagonal. This may lead to flavour changing
neutral current (FCNC) effects in the low energy effective $N=1$
softly broken Lagrangian.

A third topic of interest is the $B$-parameter, the soft mass term
which is associated to a SUSY mass term $\mu H_1H_2$ for the
pair of Higgsses $H_{1,2}$ in the Minimal Supersymmetric Standard Model (MSSM).
Compared to the other soft
terms, the result for the $B$-parameter is more model-dependent.
Indeed, it depends not only on the dilaton/moduli
dominance assumption but also on the particular mechanism which could
generate the associated ``$\mu$-term''. An interesting possibility to
generate such a term is the one suggested in ref.\cite{GM}
in which it was
pointed out that in the presence of certain bilinear terms in the
K\"ahler potential an effective $\mu$-term of order the gravitino
mass, $m_{3/2}$, is naturally
generated. Interestingly enough, such bilinear terms in the
K\"ahler potential do appear in string models and particularly in	
Abelian orbifolds. In section 4 we compute the $\mu $ and
$B$ parameters as well as the soft scalar masses of the
charged fields which could play the role of Higgs particles in
such Abelian orbifold schemes. We find the interesting result that,
independently of the Goldstino direction in the
dilaton/moduli space, one gets the prediction $|tg\beta |=1$
at the string scale. In other words, the direction
$\langle H_1 \rangle =\langle H_2 \rangle $ remains flat
{\it even after SUSY-breaking}.
The results for $B$ corresponding to other sources
for the $\mu$-term are also presented in the multimoduli case
under consideration. In particular, the possibility of generating a small
$\mu$-term from the superpotential \cite{CM} is studied.
We leave some final comments and conclusions for section 5.

\section{ Soft terms: the multimoduli case}

We are going to consider $N=1$ SUSY 4-D strings with
$m$ moduli $T_i$, $i=1,..,m$. Such notation refers to both $T$-type
and $U$-type (K\"ahler class and complex structure in the Calabi-Yau
language) fields.
In addition there will be charged matter fields $C_{\alpha }$ and the
complex dilaton field $S$.
In general we will be considering $(0,2)$ compactifications and thus the
charged
fields do not need to correspond to $27$s of $E_6$.

Before further specifying the class of theories that we are going to consider
a comment about the total number of moduli is in order.
We are used to think of large numbers of $T$ and $U$-like moduli
due to the fact that in $(2,2)$ ($E_6$) compactifications there is a
 one to one correspondence between moduli and charged
fields. However, in the case of $(0,2)$ models
with arbitrary gauge group (which is the case of
phenomenological interest) the number of moduli is drastically reduced.
 For example,
in the standard $(2,2)$ $Z_3$ orbifold there are 36 moduli $T_i$,
9 associated to the untwisted sector and 27 to the fixed points of the
orbifold.
In the thousands of $(0,2)$ $Z_3$ orbifolds one can construct by adding
different
gauge backgrounds or doing different gauge embeddings, only the
9 untwisted moduli remain in the spectrum. The same applies to models 
with $U$-fields. This is also the case for compactifications using
$(2,2)$ minimal superconformal models. Here all singlets associated
to twisted sectors are projected out when proceeding to 
$(0,2)$ \cite{Greene}. 
So, as these examples
show, in the case of $(0,2)$ compactifications
the number of moduli is drastically reduced to a few fields.
In the case of generic Abelian orbifolds one is in fact left with
only three T-type moduli $T_i$ ($i=1,2,3$), the only exceptions being
$Z_3$, $Z_4$ and $Z'_6$, where such number is 9, 5 and 5 respectively.
The number of $U$-type fields in these $(0,2)$ orbifolds oscillates
between $0$ and $3$, depending on the specific example.
Specifically, $(0,2)$ $Z_2\times Z_2$ orbifolds have 3 $U$ fields,
the orbifolds of type $Z_4,Z_6$,$Z_8,Z_2\times Z_4$,$Z_2\times Z_6$ and
$Z_{12}'$ have just one $U$ field and the rest have no untwisted $U$-fields.
Thus, apart from the three exceptions mentioned above,
this class of models has at most 6 moduli, three of $T$-type (always
present) and at most three of $U$-type. In the case of models obtained
from Calabi-Yau type of compactifications
a similar effect is expected and  only one $T$-field associated to the
overall modulus is guaranteed to exist in $(0,2)$ models.

We will consider effective $N=1$ supergravity (SUGRA)
K\"ahler potentials of the
type:
\begin{eqnarray}
& K(S,S^*,T_i,T_i^*,C_{\alpha},C_{\alpha}^*)\ = \
-\log(S+S^*)\ +\ {\hat K}(T_i,T_i^*)\ +\
{\tilde K}_{{\overline{\alpha }}{ \beta }}(T_i,T_i^*){C^*}^{\overline {\alpha}}
C^{\beta }\  & \nonumber\\
&+ \ (\
Z_{{\alpha }{ \beta }}(T_i,T_i^*){C}^{\alpha}
C^{\beta }\ +\ h.c. \ ) \ . &
\label{kahl}
\end{eqnarray}
The first piece is the usual term corresponding to the complex
dilaton $S$ which is present for any compactification whereas the
second is the K\"ahler potential of the moduli fields, where we recall
that we are denoting
the $T$- and $U$-type moduli collectively by $T_i$.
The greek indices label the matter fields and their
kinetic term functions are  given by
${\tilde K_{{\overline{\alpha }}{ \beta }}}$ and $Z_{{\alpha }{\beta }}$
to lowest order in the matter fields. The last piece is often forbidden
by gauge invariance in specific models although it may be relevant
in some cases as discussed in section 4.
In this section we are going to consider the case of diagonal metric
both for the moduli and the matter fields and leave the off-diagonal
case for the next section. Then ${\hat K}(T_i,T_i^*)$ will be a sum
of contributions (one for each $T_i$), whereas
${\tilde K_{{\overline{\alpha }}{ \beta }}}$ will be taken of the
diagonal form ${\tilde K_{{\overline{\alpha }}{ \beta }}}
\equiv \delta _{{\overline{\alpha }}{ \beta }} {\tilde K_{\alpha }}$.
The complete $N=1$ SUGRA Lagrangian is determined by
the K\"ahler potential $K({\phi }_M ,\phi^*_M)$, the superpotential
$W({\phi }_M)$ and the gauge kinetic functions
$f_a({\phi }_M)$, where $\phi_M$ generically denotes the chiral fields
$S,T_i,C_{\alpha }$. As is well known, $K$ and $W$ appear in the
Lagrangian only in the combination $G=K+\log|W|^2$. In particular,
the (F-part of the) scalar potential is given by
\begin{equation}
V(\phi _M, \phi ^*_M)\ =\
e^{G} \left( G_M{K}^{M{\bar N}} G_{\bar N}\ -\ 3\right) \ ,
\label{pot}
\end{equation}
where $G_M \equiv \partial_M G \equiv \partial G/ \partial \phi_M$
and $K^{M{\bar N}}$ is the inverse of the K\"ahler metric
$K_{{\bar N }M}\equiv{\partial}_{\bar N}{\partial }_M K$.

The crucial assumption  now is  to locate the origin of SUSY-breaking in the
dilaton/moduli sector. It is perfectly conceivable that other fields in the
theory, like charged matter fields, could contribute in a leading manner to
SUSY-breaking.
If that is the case, the structure of soft SUSY-breaking terms will be
totally
model-dependent and we would be able to make no model-independent statements at
all
about soft terms. On the contrary, assuming the seed of SUSY-breaking
originates
in the dilaton-moduli sectors will enable us to extract some interesting
results.
We will thus make
that assumption without any further justification.
Let us take the following parametrization
for the vev's of the dilaton and moduli auxiliary
fields $F^S=e^{G/2} G_{ {\bar{S}} S}^{-1} G_{\bar{S}}$ and
$F^i=e^{G/2} G_{ {\bar{i}} i}^{-1} G_{\bar{i}}$:
\begin{equation}
G_{ {\bar{S}} S}^{1/2} F^S\ =\ \sqrt{3}m_{3/2}\sin\theta e^{-i\gamma _S}\ \ ;\ \
G_{ {\bar{i}} i}^{1/2} F^i\ =\ \sqrt{3}m_{3/2}\cos\theta\ e^{-i\gamma _i}
\Theta _i \ \ ,
\label{auxi}
\end{equation}
where $\sum _i \Theta _i^2=1$ and $e^G=m^2_{3/2}$ is the gravitino
mass-squared.
The angle $\theta $ and the $\Theta _i$ just parametrize the
direction of the goldstino in the $S,T_i$ field space.
 We have also allowed for the possibility of
some complex phases $\gamma _S, \gamma _i$ which could be relevant
for the CP structure of the theory. This parametrization has the virtue that
when we plug it in the general form of the SUGRA scalar potential
eq.(\ref{pot}), its vev (the cosmological constant) vanishes by
construction. Notice that such a phenomenological approach allows us
to `reabsorb' (or circumvent) our ignorance about the (nonperturbative)
$S$- and $T_i$- dependent part of the superpotential, which is
responsible for SUSY-breaking.
It is now a straightforward
exercise
to compute the bosonic soft SUSY-breaking terms in this class of theories.
Plugging
eqs.(\ref{auxi}) and (\ref{kahl}) into eq.(\ref{pot})
one finds the following results (we recall that we
are considering here a diagonal metric for the matter fields):
\begin{eqnarray}
 & m_{\alpha }^2 = \  m_{3/2}^2 \ \left[ 1\ -\ 3\cos^2\theta \
({\hat K}_{ {\overline i} i})^{-1/2} {\Theta }_i e^{i\gamma _i}
(\log{\tilde K}_{\alpha })_{ {\overline i} j}
({\hat K}_{ {\overline j} j})^{-1/2} {\Theta }_j e^{-i\gamma _j} \ \right] \ ,
&
\nonumber \\
 & A_{\alpha \beta \gamma } =
 \   -\sqrt{3} m_{3/2}\ \left[ e^{-i{\gamma }_S} \sin\theta \right.
& \nonumber \\
& \left. - \ e^{-i{\gamma }_i} \cos\theta \  \Theta_i
({\hat K}_{ {\overline i} i})^{-1/2}
\ \left(  {\hat K}_i - \sum_{\delta=\alpha,\beta,\gamma}
(\log {\tilde K}_{\delta })_i
+ (\log h_{\alpha \beta \gamma } )_i \ \right)
\  \right] \ . &
\label{soft}
\end{eqnarray}
The above scalar masses and trilinear scalar couplings correspond
to charged fields which have already been canonically normalized.
Here $h_{\alpha \beta \gamma }$ is a renormalizable
Yukawa coupling involving three charged chiral fields and
$A_{\alpha \beta \gamma }$ is its corresponding trilinear soft term.

Physical gaugino masses $M_a$ for the canonically normalized gaugino fields
are given by $M_a=\frac{1}{2}(Re f_a)^{-1}e^{G/2}{f_a}_M
{K}^{M{\bar N}} G_{\bar N}$.
Since the tree-level gauge kinetic function is given for any 4-D string by
$f_a=k_aS$, where $k_a$ is the Kac-Moody level of the gauge factor,
the result for tree-level gaugino masses is independent of the
moduli sector and is simply given by:
\begin{equation}
M\equiv M_a\ =\ m_{3/2}\sqrt{3} \sin\theta e^{-i\gamma _S} \ .
\label{gaugin}
\end{equation}

As we mentioned above, the parametrization of the auxiliary field vev's
was chosen in such a way to guarantee the automatic vanishing of
the vev of the scalar potential ($V_0=0$). If the value of $V_0$
is not assumed to be zero
the above formulae are modified in the following simple way.
One just has to replace $m_{3/2}\rightarrow Cm_{3/2}$,
where $|C|^2=1+V_0/3m_{3/2}^2$. In addition, the formula for $m_{\alpha }^2$
gets an additional contribution given by $2m_{3/2}^2(|C|^2-1)=2V_0/3$.

The soft term formulae above are in general valid for any compactification
as long we are considering diagonal metrics. In addition one is tacitally
assuming that the tree-level K\"ahler potential and $f_a$-functions
constitute a good aproximation.
The K\"ahler potentials
for the moduli are in general  complicated functions.
To illustrate some general features of the multimoduli case
 we will concentrate here on the case of generic $(0,2)$ symmetric
Abelian orbifolds. As we mentioned above, this class of models
contains three $T$-type  moduli and (at most) three $U$-type moduli.
We will denote them collectively by $T_i$, where e.g. $T_i=U_{i-3}$; $i=4,5,6$.
For this  class of models the K\"ahler potential has the form \cite{potential}
\begin{equation}
K(\phi,\phi^*)\ =\ -\log(S+S^*)\ -\ \sum _i \log(T_i+T_i^*)\ +\
\sum _{\alpha } |C_{\alpha }|^2 \Pi_i(T_i+T_i^*)^{n_{\alpha }^i} \ .
\label{orbi}
\end{equation}
Here $n_{\alpha }^i$ are fractional numbers usually called ``modular weights"
of the matter fields $C_{\alpha }$. For each given Abelian orbifold,
independently of the gauge group or particle content, the possible
values of the modular weights are very restricted. For a classification of
modular weights for all Abelian orbifolds see ref.\cite{IL}.
Using the particular form (\ref{orbi}) of the K\"ahler potential and
eqs.(\ref{soft},\ref{gaugin}) we obtain
the following results\footnote{This analysis was also carried out, for the
particular case of the three diagonal moduli $T_i$,
in refs.\cite{japoneses} and \cite{BC},
in order to obtain unification of gauge coupling constants and to analyze 
FCNC constraints, respectively.
Some particular multimoduli examples were also considered in
ref.\cite{FKZ} .}
for the scalar masses, gaugino masses and soft trilinear
couplings:
\begin{eqnarray}
   &m_{\alpha }^2 =  \  m_{3/2}^2(1\ +\ 3\cos^2\theta\ {\vec {n_{\alpha }}}.
{\vec {\Theta ^2}}) \ , &
\nonumber\\
&  M = \  \sqrt{3}m_{3/2}\sin\theta e^{-i{\gamma }_S} \ , &
\nonumber\\
 & A_{\alpha \beta \gamma } = \   -\sqrt{3} m_{3/2}\ ( \sin\theta e^{-i{\gamma
}_S}
\ +\ \cos\theta \sum _{i=1}^6 e^{-i\gamma _i}    {\Theta }^i {\omega
}^i_{\alpha
\beta \gamma } ) \ , &
\label{masorbi}
\end{eqnarray}
where we have defined :
\begin{equation}
{\omega }^i_{\alpha \beta \gamma }\ =\ (1+n^i_{\alpha }+n^i_{\beta
}+n^i_{\gamma
}-
 {Y}^i_{\alpha \beta \gamma }    )\ \ ;\ \
{Y}^i_{\alpha \beta \gamma } \
= \ {{h^i_{\alpha \beta \gamma }}\over {h_{\alpha \beta \gamma
}}} 2ReT_i \ .
\label{formu}
\end{equation}
Notice that neither the scalar nor the gaugino masses have any explicit
dependence on $S$ or $T_i$, they only depend on the gravitino mass and
the goldstino angles.
This is one of the advantages of a parametrization in terms of such angles.
In the case of the $A$-parameter an explicit $T_i$-dependence may appear in
the term proportional to $Y^i_{\alpha \beta \gamma }$. This explicit
dependence
disappears in three interesting cases: 1) In the dilaton-dominated case
($\cos\theta =0$).
2) When the Yukawa couplings involve only untwisted ({\bf U}) particles,
i.e couplings of the type {\bf UUU}, in which case
the coupling is a constant.
3) When the particles involved in the coupling have all
overall modular weight $n_{\alpha }=-1$ (again, the coupling is
constant). This is possible for couplings of the type
${\bf U}{\bf T}_{-1}{\bf T}_{-1}$, ${\bf T}_{-1}{\bf T}_{-1}{\bf T}_{-1}$,
where the subindex indicates the
value of the overall modular weight of the twisted ({\bf T}) particle
(see below).
This is for example the case of any $Z_2\times Z_2$ orbifold.
There is a fourth case in which the $Y^i_{\alpha \beta \gamma }$-term
does not disappear but is suppressed
for large radii. This happens when the coupling $h_{\alpha \beta \gamma}$
links twisted fields, {\bf TTT}, associated to the same fixed point.
In this case one has $h_{\alpha \beta \gamma}\simeq (constant + O(e^{-T}))$
\cite{FCM} and then $Y^i_{\alpha \beta \gamma }\rightarrow 0$. In all the first
three  cases
discussed above the soft terms obtained are independent
of the values of $S$ and $T_i$.

It is appropriate at this point to recall some information about the
``modular weights'' $n_{\alpha }^i$ appearing in these expressions.
For particles belonging to the untwisted sectors one has
\begin{equation}
n_{\alpha }^i \ =\ -\delta ^i_{\alpha }\ ; \ i =1,2,3 ;\ \
n_{\alpha }^i \ =\ -\delta ^{i-3}_{\alpha }\ ; \ i = 4,5,6 \ .
\label{modu}
\end{equation}
Here $i=1,2,3$ labels the three $T$-type moduli and $i=4,5,6$ the three
(maximum) $U$-type moduli, whereas $\alpha =1,2,3$ labels the three
untwisted sectors of the orbifold.
 Each twisted sector is associated to an order $N$ twist vector
${\vec v}=(v^1,v^2,v^3)$ defined so that
$0\leq v^i< 1$, $\sum_{i=1}^3v^i=1$. In terms of the $v_i$ one finds
the following modular weights for particles in twisted sectors:
\begin{eqnarray}
& n_{\alpha }^i\ =\ -(1-v^i+p^i-q^i)\ ;\ i=1,2,3 ;\ \ (v^i\not= 0) \ , &
\nonumber\\
&  n_{\alpha }^{i+3}\ =\ -(1-v^i+q^i-p^i)\ ;\ i=1,2,3; \ \  (v^i\not=0) \ , &
\nonumber\\
& n_{\alpha }^i=n_{\alpha }^{i+3}\ =\ 0 \ \ (v^i=0) \ , &
\label{modt}
\end{eqnarray}
where $p^i$ and $q^i$ denote the number of (left-handed) oscillator operators
of each chirality in the $i$-th complex direction (see ref.\cite{IL}
for details).
The ``overall T modular weights'' corresponding to the ``overall modulus''
$T$ field considered in ref.\cite{BIM} are given by
$n_{\alpha }=\sum _{i=1}^3 n_{\alpha }^i$.
Twisted sectors with all $v^i\not=0$ (and no oscillators)
have overall modular weights $n_{\alpha }=-2$
due to the property $\sum _{i=1}^3v^i=1$. Twisted sectors with one of the
$v^i$ vanishing have the form ${\vec v}=(1/r,(r-1)/r,0)$
(plus permutations) with $r=2,3,4,6$. Such sectors obviously have
overall modular weights
$n_{\alpha }=-1$. If the twisted particle has also $p$ ($q$)
positive (negative) chirality oscillators, the overall $T$ modular weight
gets an extra addition $=p-q$. Particles with oscillators normally correspond
to small representations of the gauge group (e.g., singlets) so that
one expects the interesting charged particles to be associated
to either untwisted sector or twisted sectors with no oscillators
(or perhaps at most one or two oscillators).

With the above information we can now analyze the different structure of
soft terms available for each Abelian orbifold. The results obtained in
ref.\cite{BIM}
corresponded to the assumption that only $S$ and the overall modulus
$T$ were the seed of SUSY breaking. Within the more general framework here
described, those results correspond to
the particular goldstino direction
\begin{equation}
{\vec {\Theta ^2}}\ =\ ({1\over 3},{1\over 3},{1\over 3},0,0,0)\ \
\label{era}
\end{equation}
and can be recovered from eq.(\ref{masorbi}) and eq.(\ref{formu}) (assuming
also $\gamma_i=\gamma_T$,
$h^i_{\alpha \beta \gamma}=h^T_{\alpha \beta \gamma}/3$):
\begin{eqnarray}
   &m_{\alpha }^2 =  \  m_{3/2}^2(1\ +\ n_{\alpha}\cos^2\theta) \ , &
\nonumber\\
&  M = \  \sqrt{3}m_{3/2}\sin\theta e^{-i{\gamma }_S} \ , &
\nonumber\\
 & A_{\alpha \beta \gamma } = \   -\sqrt{3} m_{3/2}\ ( \sin\theta e^{-i{\gamma
}_S}
\ +\ \frac{1}{\sqrt{3}}\cos\theta e^{-i\gamma_T}
{\omega }_{\alpha \beta \gamma } ) \ , &
\label{masorbio}
\end{eqnarray}
where we have defined :
\begin{equation}
{\omega }_{\alpha \beta \gamma }\ =\ (3+n_{\alpha}+n_{\beta}+n_{\gamma}-
 {Y}^T_{\alpha \beta \gamma }    )\ \ ;\ \
{Y}^T_{\alpha \beta \gamma } \
= \ 2ReT {{h^T_{\alpha \beta \gamma }}\over {h_{\alpha \beta \gamma
}}} \ .
\label{formuo}
\end{equation}
In that case one could extract a number of generic qualitative properties of
soft terms  with regard to three important issues : the existence or not
of negative mass$^2$ for some matter fields, the universality of soft
scalar masses,
and the relative sizes of
gaugino versus scalar masses. In the case of an overall $T$ modulus
one finds (see the above formulae):

{\it 1)} Scalars in untwisted and in twisted sectors with overall $T$-modular
weight $n_{\alpha } =-1$
have always masses-squared $\geq 0$.

{\it 2)} Scalars in twisted sectors with $n_{\alpha }\leq -2 $ are always
lighter than those with $n_{\alpha }=-1$. The condition
$\cos^2\theta \leq 1/|n_{\alpha }|$ is required for a particle $C_{\alpha }$
not to become tachyonic.

{\it 3)} Universal soft scalar masses are obtained in two cases: First, in the
dilaton-dominated SUSY-breaking ($\cos\theta =0$) which implies that the
whole soft terms are universal (see eq.(\ref{masorbi}))
\cite{KL,BIM}. Second, if all scalars
have the same overall modular weight $n_{\alpha}=n$ \cite{BIM}.
For example, this always occurs for any $Z_2\times Z_2$ orbifold.

{\it 4)} Due to the above constraints, all scalars $C_{\alpha }$ verify
$M^2\geq m^2_{\alpha }$.

We would like now to study to what extent these general conclusions change
in the multimoduli case. We will discuss them in turn.

{\it 1) Soft masses for  $n_{\alpha }=-1 $ particles}

Let us start with the first of these issues, the masses of
$n_{\alpha }=-1$ sectors. There are two types of such sectors, the untwisted
sector (which is present in any orbifold) and the twisted sectors
with $n_{\alpha }=-1$.
We will discuss them in turn. Using the formulae above one finds
the following expressions for scalars in the three untwisted sectors of any
orbifold:
\begin{eqnarray}
& m_1^2\ =\ m_{3/2}^2\ (1-3\cos^2\theta (\Theta ^2_1+\Theta ^2_4)  ) \ , &
\nonumber\\
& m_2^2\ =\ m_{3/2}^2\ (1-3\cos^2\theta (\Theta ^2_2+\Theta ^2_5)  ) \ , &
\nonumber\\
& m_3^2\ =\ m_{3/2}^2\ (1-3\cos^2\theta (\Theta ^2_3+\Theta ^2_6)  ) \ . &
\label{untw}
\end{eqnarray}
One immediately observes that the only way to avoid the presence of
tachyons for {\it any} choice of goldstino direction
in all three sectors is imposing the condition $\cos^2\theta \leq 1/3$.
This is to be compared to the overall modulus case (\ref{masorbio}) in which
positive mass$^2$ was obtained for any $\theta $.
Notice the following important sum-rule which is valid
for the untwisted particles of any orbifold:
\begin{equation}
m_1^2\ +\ m_2^2\ +\ m_3^2\ =\ |M|^2\ \ .
\label{rulix}
\end{equation}
Furthermore, since ${\vec {n_1}}+{\vec {n_2}}+{\vec {n_3}}=-(1,1,1,1,1,1)$
and the {\bf UUU}
Yukawa couplings do not depend on the moduli
one also has
\begin{equation}
A_{123}\ =\ -M \ .
\label{aaa}
\end{equation}

Let us consider now the case of twisted sectors with $n_{\alpha }=-1$.
As we said, the associated twist vectors have the form
${\vec v}=(1/r,(r-1)/r,0)$ (plus permutations) with $r=2,3,4,6$.
Looking at the first of the eqs.(\ref{masorbi})\  one sees that
one has guaranteed a positive mass$^2$ if
$\cos^2\theta \leq r/3(r-1)$. The tighter bound is obtained when $r=6$
which yields $\cos^2\theta \leq 2/5$. A generalization of eqs.(\ref{rulix})
and (\ref{aaa})\ apply also in this case. Consider three particles
$C_{\alpha }$,$C_{\beta }$,$C_{\gamma }$ all with overall modular weight
$=-1$ coupling through a Yukawa $h_{\alpha \beta \gamma }$. They may belong
both
to the untwisted sector or to a twisted sector with $n=-1$, i.e. couplings
of the type ${\bf U}{\bf T}_{-1}{\bf T}_{-1}$,
${\bf T}_{-1}{\bf T}_{-1}{\bf T}_{-1}$.
Then it is easy to
convince oneself that again for any possible twist
${\vec {n_\alpha }}+{\vec {n_\beta }}+{\vec {n_\gamma }}=-(1,1,1,1,1,1)$.
Then one
finds
that for {\it any choice} of goldstino direction
\begin{equation}
m_{\alpha }^2\ +\ m_{\beta }^2\ +\ m_{\gamma }^2\ =\ |M|^2\
=3 m_{3/2}^2\sin^2\theta \
\label{rulox}
\end{equation}
and besides
\begin{equation}
A_{\alpha \beta \gamma }\ =\ -M \ .
\label{ruloxxt}
\end{equation}
The only difference with eqs.(\ref{rulix}), (\ref{aaa}) is that
eqs.(\ref{rulox}), (\ref{ruloxxt}) apply to
any three $n=-1$  particles linked by a Yukawa coupling
(and not only to the three untwisted sectors). Thus, for example, the sum-rule
applies to any set of three particles which couple in any $Z_2\times Z_2$
orbifold.
Specific examples will be shown below.

Notice that if we insist in having a vanishing gaugino mass, the sum-rules
(\ref{rulix}) and (\ref{rulox}) force
the  scalars to be either all massless or at least one of them tachyonic.
As we will discuss below, having a tachyonic sector is not necessarily a
problem, it may even be an advantage, so one should not disregard  this
possibility
at this point. Of course, in the trivial case when there is no physical
particle in that particular sector which would have negative mass$^2$
the situation is also harmless. Let us show an explicit example of this
possibility.
Consider the second example of Table 3 of ref.\cite{INQ}. This is a
three-generation
$Z_3$
orbifold model with gauge group $SU(3)_c\times SU(3)_L\times SU(3)_R$.
It has the particular property that it has no charged matter in the
untwisted sector so that the sum-rule (\ref{rulix}) can cause no trouble in the
untwisted sector (i.e., no physical tachyons). Consider the goldstino
direction e.g. ${\vec {\Theta }}=(0,0,1)$. The untwisted particles would have
had masses
$m_1^2=m_2^2=m_{3/2}^2$, $m_3^2=m_{3/2}^2(1-3\cos^2\theta )$ whereas the
twisted particles would have
$m_{\bf T}^2=m_{3/2}^2(1-2\cos^2\theta )$. The absence of
 charged massless particles in the untwisted sector would have allowed us  to
have e.g., $1/3\leq \cos^2\theta \leq 1/2$, values which would have lead to
tachyonic
states in the untwisted sector. For the particular value $\cos^2\theta =1/2$
one gets $m_{\bf T}^2=0$ and gaugino masses $M^2=3/2m^2_{3/2}$.

From the above discussion we conclude that in the
multimoduli case, depending on the goldstino direction, tachyons
may appear both in the untwisted and $n_{\alpha} =-1$ twisted sectors unless
$\cos^2\theta \leq 1/3$. This is to be compared to the overall modulus $T$ case
in which
tachyons never appear.  For $\cos^2\theta \geq 1/3 $, one has to
be very careful with the goldstino direction if one is interested
in avoiding tachyons. In some sense, a certain amount of fine tuning
is required so that the goldstino direction goes more and more in the
overall $T$ modulus direction as one increases $\cos^2\theta $.
Nevertheless we should not forget that tachyons, as we already mentioned above,
are not necessarily a problem, but may just show us an instability.

{\it 2) Soft masses for $n_{\alpha }= -2 $ particles }

In the absence of oscillators,
these are particles originated in twisted sectors
${\vec v}=(v^1,v^2,v^3)$ with all $v^i\not=0$. Plugging the expressions for the
modular weights one finds in this case
\begin{equation}
  m_{\alpha }^2 =\  m_{3/2}^2(1 - 3\cos^2\theta )\ +\
 3m_{3/2}^2\cos^2\theta {\vec v}_{\alpha }.{\vec {\Theta ^2}} \ ,
\label{masorba}
\end{equation}
where ${\vec v}_{\alpha }=(v^1,v^2,v^3,v^1,v^2,v^3)$.
It is obvious from
eq.(\ref{masorba}) that having $\cos^2\theta \leq 1/3$ will be enough to
guarantee the
absence of tachyons for any $n=-2$ particle. This is to be compared with
the overall modulus case analyzed in ref.\cite{BIM}
in which the weaker condition
$\cos^2\theta \leq 1/2$ was required. Notice also that in the overall modulus
$T$ case one always had that the $n=-1$ scalar had bigger masses than the
$n=-2$ scalars. Here the situation may even be reversed.
For any three  fields $C_{\alpha}$,$C_{\beta }$,$C_{\gamma}$
linked through a ${\bf T}_{-2}{\bf T}_{-2}{\bf T}_{-2}$ Yukawa coupling
one can check the following sum-rule which is true for any goldstino direction
${\vec {\Theta }}$ :
\begin{equation}
m_{\alpha }^2+m_{\beta }^2+m_{\gamma }^2\ =\ 3m_{3/2}^2(1-2\cos^2\theta)
\ =\ |M|^2\ -\ 3m_{3/2}^2\cos^2\theta \ .
\label{rulax}
\end{equation}
This shows us that, {\it on average}, $n=-2$ twisted particles are lighter than
$n=-1$ particles but the reverse may be true for some particular fields
as long as the above sum-rules are not violated.

It is worth noticing here that twisted Yukawa couplings mixing particles with
$n=-1$ and $n=-2$ are also possible
(e.g. ${\bf T}_{-1}{\bf T}_{-2}{\bf T}_{-2}$,
${\bf T}_{-1}{\bf T}_{-1}{\bf T}_{-2}$). In this case the sum-rule is
\begin{equation}
m_{\alpha }^2+m_{\beta }^2+m_{\gamma }^2\ =\ |M|^2\ -
\ 3m_{3/2}^2\cos^2\theta\ \delta
\label{rulaxxt}
\end{equation}
with
\begin{equation}
\delta\equiv 1- \sum_{k} \Theta_k^2 \ ,
\label{rulaxxtt}
\end{equation}
where $\Theta_k$ are the auxiliary fields of the moduli associated to the
vanishing entry of the $n=-1$ twist vectors (see below eq.(\ref{aaa})) present
in the coupling, i.e. those with $n_{\alpha}^k=0$. Since $0< \delta <1$, the
sum-rule (\ref{rulaxxt}) is rather in-between the (\ref{rulox}) and the
(\ref{rulax}).

Let us finally comment that if the twisted particle has associated an
oscillator
operator, the modular weight decreases in as many units as (positive
chirality) oscillators. This makes very likely for such particles to
have negative mass$^2$ (unless there is approximate dilaton dominance)
. In many  cases such particles are just singlets
and  such tachyonic behaviour may just denote that these fields are forced
to aquire vev's.

{\it 3) Universality of soft scalar masses}

In the dilaton-dominated case ($\cos\theta =0$) the whole
soft terms are universal
as in the overall modulus case. Also scalars with different overall
modular weights
$n_{\alpha}$ have different masses.
However, unlike the overall modulus case,
non-universal soft scalar masses for particles with the same
$n_{\alpha}$ are allowed and in fact this will be the most general situation
(see e.g. eqs.(\ref{untw},\ref{masorba})).

{\it 4) Gaugino versus scalar masses}

In the overall modulus $T$ discussed in ref.\cite{BIM}
the heaviest scalars were the
ones
with
modular weight $n=-1$ which had mass$^2=|M|^2/3$. So scalars are lighter than
gauginos
at this level. In the multimoduli case sum-rules like (\ref{rulox})  replace
the equation $3m^2_{n=-1}=|M|^2$. In some way, on average the scalars
are lighter than
gauginos but there may be scalars with mass bigger than gauginos. In the case
of
particles with $n=-1$, eq.(\ref{rulox})\ tells us that this can only be true
at
the cost of
having some of the other three scalars with {\it negative} mass$^2$.
This may have diverse phenomenological
implications depending what is the particle content
of the model, as we now explain in some detail:

{\it 4-a) Gaugino versus scalar masses in standard model 4-D strings}

Let us consider first the case of string models with gauge group
$SU(3)_c\times SU(2)_L\times U(1)_Y$$\times G$ and see whether
one can avoid the general situation of ref.\cite{BIM}, where scalar
masses were found to be always smaller than gaugino masses (at tree-level).
In the present more general framework, one can certainly find explicit 
examples of orbifold sectors where some individual scalar mass is bigger 
than gaugino masses even at the tree-level. For example, let us consider 
the case of the $Z_8$ orbifold with an observable particle in the twisted sector
${\bf T}_{\theta^6}$. The modular weight associated to that sector is
${\vec {n_{\theta^6}}}=(-1/4,-3/4,0,0)$ and therefore (see eq.(\ref{masorbi}))
\begin{equation}
m_{\theta^6}^2\ =\ m_{3/2}^2\ \left[1-3\cos^2\theta
\left(\frac{1}{4}\Theta^2_1+\frac{3}{4}\Theta^2_2\right) \right] \ .
\label{masa}
\end{equation}
Then, choosing e.g. a goldstino direction with $\cos^2\theta=5/6$, 
$\Theta_1=\Theta_2=0$, one gets $m_{\theta^6}^2=m_{3/2}^2$, $M^2=m_{3/2}^2/2$. 
Many more examples along these lines can be found of course. In   
general one finds that it is possible to get $m_{\alpha} > M$, 
provided $\sin\theta$ is sufficiently small. Indeed, from the general 
formulae eq.(\ref{masorbi}) we see that always $m_{\alpha}\leq m_{3/2}$ and 
therefore a necessary (although usually not sufficient) condition to get 
scalars heavier than gauginos is
\begin{equation}
\cos^2\theta > 2/3 \; .
\label{coseno}
\end{equation}

After such preliminary remark one immediately realizes that, 
especially in the case of standard model 4-D strings, 
further important restrictions on the possibility of getting 
scalars heavier than gauginos come from sum-rules like
(\ref{rulix},\ref{rulox},\ref{rulax},\ref{rulaxxt}), 
which typically constrain the masses of three particles linked
via a Yukawa coupling. Suppose that all the three particles involved
are observable particles (squarks, sleptons, Higgses). 
If we require that the corresponding squared masses be non-negative 
in order to avoid automatically phenomenological problems such 
as charge and color breaking or Planck scale Higgs vevs, 
then the sum rule will immediately imply that such masses
are smaller than gaugino masses. Conversely, if we tried to
obtain one scalar mass bigger than gaugino masses by an
appropriate choice of the goldstino direction, then at least 
one of the other two scalar masses would become tachyonic.  
On the other hand, tachyons may be helpful if the particular Yukawa coupling
does not involve observable particles. They could break extra gauge symmetries
and generate large masses for extra particles. We recall that standard-like
models in strings usually have too many extra particles and many extra
U(1) interactions. Although the Fayet-Iliopoulos mechanism helps to cure
the problem \cite{suplemento}, the existence of tachyons is a complementary
solution.

Concerning observable particles, we have just seen that the sum
rules, supplemented by `no-tachyon' requirements, typically lead 
to the conclusion that observable scalars are lighter than gauginos
\begin{equation}
m_{\alpha} < M \ ,
\label{masas1}
\end{equation}
similarly to the situation found in the symplified scenario of 
ref.\cite{BIM}. Therefore, since gaugino loops play a main
role in the renormalization of scalar masses down to low-energy,
the gluino, slepton and (first and second generation) squark mass 
relations at the electroweak scale turn out (again) to be
\begin{equation}
m_l < m_q \simeq M_g \ ,
\label{masas2}
\end{equation}
where gluinos are slightly heavier than squarks. We recall that
slepton masses are smaller than squark masses because they do not 
feel the important gluino contribution.

It is still possible to ask whether the generic situation
described by eqs.(\ref{masas1}) and (\ref{masas2}) admits
exceptions. 
One possibility
is the following.
One could get some squark or slepton mass bigger than gaugino masses
by allowing a negative soft squared mass for a Higgs field, provided
the total squared Higgs mass (including the $\mu^2$ contribution) is 
non-negative\footnote{Notice that such a possibility can be explored in 
detail only after specifying the mechanism for generating the $\mu$ parameter 
itself (see e.g. ref.\cite{Nuestra}).}.
Another possibility which comes to mind is the case
in which a Yukawa coupling among `observable' particles
originates actually from a non-renormalizable (rather than 
renormalizable) coupling\footnote{Notice however that
this is unlikely to be the case for the top Yukawa coupling,
which is relevant e.g. for radiative symmetry breaking.}, 
where the extra fields in the coupling get vevs (e.g. 
$H_2Q_Lu_L^c<\phi...\phi>$ rather than just $H_2Q_Lu_L^c$). 
In such a case new sum-rules would apply to the full set of fields in the
coupling and the above three-particle sum-rules could be violated. In
particular, observable scalars would be allowed to be heavier than gauginos,
possibly at the price of having some tachyon among the (standard model 
singlet) $\phi$ fields. In both cases
mentioned here one could get a violation of (\ref{masas1}) for some
scalars, i.e.
\begin{equation}
m_{\alpha} > M_a .
\label{masas3}
\end{equation} 
However we recall from our initial discussion that this can happen
only for small $\sin\theta$ and special goldstino directions.
Moreover, even for small (but not too small) $\sin\theta$, scalar 
and gaugino masses will be still of the same order, so that the
low-energy relation (\ref{masas2}) will still hold.
The only difference is that now squarks,
 fulfilling eq.(\ref{masas3}),
will be slightly heavier than
gluinos.
In order to reverse the situation and get instead
\begin{equation}
M_g < m_l , m_q  
\label{masas4}
\end{equation}
one needs one of the above `mechanisms' and very small $\sin\theta$,
so that $m_{\alpha} >> M_a$. Note that in such a limit
additional attention should be payed to avoid that a
too large scalar-to-gaugino mass ratio could spoil the 
solution to the gauge hierarchy problem.   

Before concluding, we recall that a pattern like (\ref{masas4})
for very small $\sin\theta$ was also obtained in the overall modulus
analysis of ref.\cite{BIM} for different reasons, i.e. as an effect of
string loop corrections to $K$ and $f_a$. After the inclusion of
such corrections the masses of gauginos and $n_{\alpha}=-1$ scalars,
which vanish at tree-level for $\sin\theta \rightarrow 0$, become
nonvanishing and typically satisfy relation (\ref{masas3}). One difference 
with the previous case is that the loop-induced case gives scalar masses 
smaller than $m_{3/2}$ instead than ${\cal O}(m_{3/2})$. In addition,
one may consider this possibility of obtaining scalars heavier than 
gauginos as a sort of fine-tuning. In the absence of a more fundamental 
theory which tells us in what direction the goldstino angles point, one 
would naively say that the most natural possibility would be to assume 
that all moduli contribute to SUSY-breaking in more or less (but {\it not} 
exactly) the same\footnote{For an explicit example of this, using gaugino 
condensation, see ref.\cite{Bailin}.} amount.

Summarizing the situation concerning standard model strings,
we have seen that the overall modulus results are qualitatively
confirmed, in the sense that for generic goldstino directions (with not
too small $\sin\theta$) the low-energy pattern of eq.(\ref{masas2})
typically holds, mainly because of the restrictions coming from 
mass sum rules and absence of tachyons. Possible exceptions giving rise
to patterns like (\ref{masas4}) may exist for special goldstino angles, 
necessarily including a sufficiently small $\sin\theta$.

{\it 4-b ) Gaugino versus scalar masses in GUT 4-D strings}

What it turned out to be a potential disaster
in the case of standard model strings may be an interesting
advantage in the case of string-GUTs.
In this case it could well be that
the negative mass$^2$ may
just induce gauge symmetry breaking by forcing a vev for a particular
scalar (GUT-Higgs field) in the model.
The latter possibility provides us with interesting phenomenological
consequences.
Here the breaking of SUSY would directly induce further gauge symmetry
breaking.

Let us now show an explicit example of the different possibilities discussed
above (scalars lighter or heavier than gauginos) in the context of GUTS.
We are going to consider a $Z_2\times Z_2$ orbifold
model
which is an $SO(10)$ string-GUT recently constructed in ref.\cite{AFIU}. We
show
in Table 1 the particle content of the model and the quantum numbers of the
particles with respect to the gauge group
$SO(10)\times (SO(8)\times U(1)^2)$. The three untwisted sectors are denoted
by ${\bf U}_1$,${\bf U}_2$,${\bf U}_3$ and the three twisted sectors by
${\bf T}_{\theta}$,${\bf T}_{\omega}$ and ${\bf T}_{\theta \omega}$.
This model has a GUT-Higgs field transforming as a $54$ of $SO(10)$
in the ${\bf U}_3$ untwisted sector. Four net generations as well as
two pairs $16+{\overline {16}}$ are present in the
${\bf T}_{\theta}$,${\bf T}_{\omega}$
twisted sectors. Finally, $10$-plets adequate to do the electro-weak
symmetry breaking belong to the ${\bf T}_{\theta \omega}$ sector.
Yukawa couplings of the following types are present in the
model:
\begin{equation}
{\bf U}_1{\bf U}_2{\bf U}_3\ \ ,\ \
{\bf U}_3{\bf T}_{\theta \omega}{\bf T}_{\theta \omega}\ \  , \ \
{\bf T}_{\theta}{\bf T}_{\omega}{\bf T}_{\theta \omega}
\label{yuk}
\end{equation}
(Not all of the latter two couplings are allowed since the
space-group selection rules may forbid some of them.)
All Yukawa couplings are constants, do not depend on $T_i$
\cite{FCM}.

The $Z_2\times Z_2$ orbifold
has three $T$ moduli and three $U$ moduli in the untwisted sector
but we are considering in this example for simplicity the case in which
only $S$ and the $T_i,i=1,2,3$ participate in SUSY-breaking. The modular
weights
of the
different sectors are:
\begin{eqnarray}
& {\vec {n_1}} =(-1,0,0)\ ;\ {\vec {n_2}}=(0,-1,0)\ ;\ {\vec {n_3}}=(0,0,-1)
\ , &
\nonumber\\
& {\vec {n_{\theta }}} =(0,-1/2,-1/2)\ ;\ {\vec {n_{\omega }}}=(-1/2,0,-1/2)\
;\
{\vec {n_{\theta \omega }}}=(-1/2,-1/2,0) \ . &
\label{modor}
\end{eqnarray}
All the sectors in the $Z_2\times Z_2$ orbifold have overall modular weight
=--1
and hence the sum-rule (\ref{rulox}) applies for any three set of particles
linked by a Yukawa coupling. Notice in particular that
${\vec {n_\alpha }}+{\vec {n_\beta }}+{\vec {n_\gamma }}=-(1,1,1,1,1,1)$
for the sets of particles related by the Yukawas (\ref{yuk}). Thus, for {\it
any goldstino
angle} one has the constraints:
\begin{eqnarray}
& m_1^2+m_2^2+m_3^2\ =\ m_\theta ^2 +m_{\omega }^2+m_{\theta \omega }^2\ =\
m_3^2 +m_{\theta \omega }^2+m_{\theta \omega }^2\ =\ M^2 \ , &
\nonumber\\
& A_{123}\ =\ A_{\theta \omega (\theta \omega )}\ =\
A_{3(\theta \omega )(\theta \omega )}\ =\ -M \ . &
\label{cons}
\end{eqnarray}
To study the different effects of chosing different goldstino directions let us
consider several examples:

\begin{table}
\begin{center}
\begin{tabular}{|c|c|c|c|c|c|c|c|}
\hline
$Sector $
& $SO(10)\times SO(8)$ & $Q$ &  $Q_A$ & $A$ & $
B$ &
$C$ & $D$
 \\
\hline
 $gauginos $ & $(45,1)+(1,28)$ & 0 & 0
& $3m_{3/2}^2$ & $m_{3/2}^2$ & $m_{3/2}^2$ & $    0  $
 \\
\hline
$  {\bf U}_1  $  &     (1,8) &   1/2   &   1/2
& $m_{3/2}^2$ & $0$  &  $m_{3/2}^2$ & $m_{3/2}^2$  \\
\hline
& (1,8)   &    -1/2   &  -1/2
& $m_{3/2}^2$ &  $0$   & $m_{3/2}^2$ & $m_{3/2}^2$
\\
\hline     $  {\bf U}_2  $ &
(1,8)   & -1/2 & 1/2   & $m_{3/2}^2$ & $0$  & $m_{3/2}^2$ & $m_{3/2}^2$
\\
\hline    & (1,8)    &
1/2  &  -1/2   & $m_{3/2}^2$ &  $0$  & $m_{3/2}^2$ &
$m_{3/2}^2$\\
\hline      $   {\bf U}_3  $ &
(54,1)   & 0   &  0   & $m_{3/2}^2$ &  $m_{3/2}^2$ & $-m_{3/2}^2$ &
$-2m_{3/2}^2$\\
\hline     & (1,1)    &
0  & 0    & $m_{3/2}^2$& $m_{3/2}^2$ &
$-m_{3/2}^2$ &$-2m_{3/2}^2$\\
\hline
 & (1,1) &  0  & 1 & $m_{3/2}^2$ & $m_{3/2}^2$ &$-m_{3/2}^2$ &
$-2m_{3/2}^2$\\
\hline   & (1,1)
&   1  & 0    & $m_{3/2}^2$ & $m_{3/2}^2 $ &$-m_{3/2}^2$ &
$-2m_{3/2}^2$\\
\hline       & (1,1)
&  -1  & 0    & $m_{3/2}^2$ &$m_{3/2}^2 $ & $-m_{3/2}^2$ &
$-2m_{3/2}^2$\\
\hline     & (1,1)
&  0  & -1    & $m_{3/2}^2$ & $m_{3/2}^2 $ &$-m_{3/2}^2$ &
$-2m_{3/2}^2$\\
\hline      ${\bf T}_{\theta}$
& $3(16,1)$   &  1/4  &  1/4   & $m_{3/2}^2$ &  $1/2m_{3/2}^2$ & 0  &
$-1/2m_{3/2}^2$\\
\hline     &
$(\overline{16},1)$   &  -1/4  &  -1/4   & $m_{3/2}^2$ & $1/2m_{3/2}^2$ &   0
&
$-1/2m_{3/2}^2$\\
\hline     ${\bf T}_{\omega}$
& $3(16,1)$   &  -1/4  &  1/4   & $m_{3/2}^2$ & $1/2m_{3/2}^2$ & 0    &
$-1/2m_{3/2}^2$\\
\hline         &
$(\overline{16},1)$   &  1/4  &  -1/4   & $m_{3/2}^2$ & $1/2m_{3/2}^2$ &  0
&
$-1/2m_{3/2}^2$\\
\hline
${\bf T}_{\theta \omega}$ &    $ 4(10,1)$   & 0   & 1/2   & $m_{3/2}^2$ & $0$ &
$m_{3/2}^2$
&
$m_{3/2}^2$\\
\hline        &
$4(10,1)$   & 0   & -1/2   & $m_{3/2}^2$ & $0$ &$m_{3/2}^2$ &
$m_{3/2}^2$\\
\hline       &
$3(1,8)$   & 0   & 1/2   & $m_{3/2}^2$ & $0$ &$m_{3/2}^2$ &
$m_{3/2}^2$\\
\hline       &
$(1,8)$   & 0   & -1/2   & $m_{3/2}^2$ & $0$ &$m_{3/2}^2$ &
$m_{3/2}^2$\\
\hline             &
$8(1,1)$   & 1/2   & 0   & $m_{3/2}^2$ & $0$ &$m_{3/2}^2$ &
$m_{3/2}^2$\\
\hline        &
$8(1,1)$   & -1/2   & 0   & $m_{3/2}^2$ & $0$ &$m_{3/2}^2$ &
$m_{3/2}^2$\\
\hline
\end{tabular}
\end{center}
\caption{Particle content and charges of the string-GUT example discussed in
the
text. The four rightmost columns desplay four examples of consistent
soft masses from dilaton/moduli SUSY breaking.}

\label{tuno}
\end{table}

A) Dilaton dominance: $\cos^2\theta =0$. All scalars
have masses $m_{\alpha }^2=m_{3/2}^2$ and
$M^2=3m_{3/2}^2$. The same universal $M/m_{\alpha }$ ratio is
mantained
in the overall modulus case (i.e., ${\vec {\Theta^2 }}=(1/3,1/3,1/3)$)
for any $\theta $. This happens because $n_{\alpha }=-1$ in
eq.(\ref{masorbio}).

B) Consider the goldstino direction  ${\vec {\Theta^2 }}=(1/2,1/2,0)$ and
$\cos^2\theta=2/3$. One finds $|M|^2=|A|^2=m_{3/2}^2$ and the scalars get
masses
as shown in column B of Table 1. The soft masses are no longer universal since
e.g. the masses of the electroweak doublets and the generations are
different. This
is important e.g. in computing electro-weak radiative symmetry breaking.

C) Consider the goldstino direction ${\vec {\Theta^2 }}=(0,0,1 )$ and
$\cos^2\theta =2/3$. One still has $|M|^2=|A|^2=m_{3/2}^2$ but now
the GUT-Higgs $54$ and the singlets get negative mass$^2$
(see column C in Table 1). This will
drive a large vev (of order the string scale) $<54>$. Although one would
naively think that the potential becomes unbounded below, one has to recall
that
the matter metrics that we are using are correct to leading order on the matter
fields
and hence for vev's of order of the string scales the potential should be
stabilized.

D) Consider finally the  direction  ${\vec {\Theta^2 }}=(0,0,1 )$ but
$\cos^2\theta =1$, i.e., only the modulus $T_3$ contributes to SUSY-breaking
(no dilaton contribution). Now the gauginos are massless, the
$10$-plets have positive masses but both the $54$ and the $16+{\overline {16}}$
pairs will tend to get vev's (see column D in Table 1).

As the above examples show, different possibilities are obtained for each given
orbifold  model depending on the particular goldstino direction. However,
not any possibility may be realized within a given class of models.
For example, the addition of any combination of soft terms violating
the constraints (\ref{cons}) would be inconsistent with the
hypothesis of dilaton/moduli induced SUSY-breaking. The reader may check that
indeed
the four choices of soft terms shown in the Table verify the constraints in
(\ref{cons}).

Comparing the conclusions of this section with those found in ref.\cite{BIM}
one certainly
finds plenty of differences. However the reader must keep in mind that e.g.
the examples B,C,D above correspond to extreme cases in which some modulus does
not
participate at all in the process of symmetry breaking. On the other hand the
overall modulus case is also in some way an extreme case since the different
moduli
participate in {\it exactly} the same way, which is also a sort of
fine-tuning.
As already mentioned above,
in the absence of a more fundamental theory which tells us in what direction
the
goldstino angles point, one would naively say that the most natural possibility
would be to assume that all moduli contribute to SUSY-breaking in more or
less (but not exactly the same) amount. In this case the conclusions would
be
half-way in-between the results found in this section and those found in
ref.\cite{BIM}.
In this context we must remark the sum-rules discussed above which would be
valid
for any choice of goldstino directions.
Let us finally remark that, in spite of the different possibilities of soft
masses in the multimoduli case, the most natural (slepton-squark-gluino)
mass relations
{\it at low-energy} will be similar to the ones of
the overall modulus case eq.(\ref{masas2}) as shown in point {\it 4-a}.

\section{ Off-diagonal matter metric}

In the previous chapter we confined ourselves to the case of diagonal
matter metric
${\tilde K_{{\overline{\alpha }}{ \beta }}}
\simeq \delta _{{\overline{\alpha }}{ \beta }}$. In fact that assumption is
justified for most of the Abelian orbifold models.
The reason is that, in the case of twisted sectors, each particle has
associated
space-group discrete quantum numbers which forbid off-diagonal metrics
(we are talking here about singular, non-smoothed out $(0,2)$ orbifolds).
In the case of matter fields in untwisted sectors, both gauge invariance and
discrete R-symmetries from the right-moving sector forbids off-diagonal
terms in almost all cases. There are only three exceptions to this general
rule,
the $(0,2)$ models based on the orbifolds $Z_3$,$Z_4$ and $Z_6'$. They are
precisely
the only Abelian orbifolds in which there are more than three $T_i$ moduli,
9, 5 and 5 respectively. They also have in common the existence of an
enhanced non-Abelian gauge symmetry in their $(2,2)$ versions ($SU(3)$ in the
first case, $SU(2)$ in the other two).
An off-diagonal metric only appears for fields in the untwisted
sectors of those examples.
In spite of the relative rareness of
off-diagonal metric in orbifolds, it is worth studying what new features
can appear in this case compared to the diagonal one, since
off-diagonal metrics could be present in other less simple
(e.g., Calabi-Yau) compactifications.

First we go back to eq.(\ref{kahl}) and compute the scalar soft
terms in the most general case where the moduli and matter metrics
are not diagonal.
Then the soft mass matrix ${\cal M}'^2_{ {\overline{\alpha }} { \beta } }$
(corresponding to unnormalized charged fields) and the soft parameters
$A_{\alpha\beta\gamma}$ read
\begin{eqnarray}
\label{mmatrix}
{\cal M}'^2_{{\overline{\alpha }}{ \beta }}  & = &
m_{3/2}^2 {\tilde K_{{\overline{\alpha }}{ \beta }}}
- {\overline F}^{\overline{i}} ( \partial_{\overline{i}}\partial_j
{\tilde K_{{\overline{\alpha }}{ \beta }}}
-\partial_{\overline{i}} {\tilde K_{{\overline{\alpha }}{ \gamma}}}
{\tilde K^{{ \gamma} {\overline{\delta}} }}
\partial_j {\tilde K_{{\overline{\delta}}{ \beta}}}  ) F^j
\\
A_{\alpha\beta\gamma} & = &
F^S K_S h_{\alpha\beta\gamma} + \delta A_{\alpha\beta\gamma}
\\
\delta A_{\alpha\beta\gamma} & = &
F^i \left[ {\hat K}_i h_{\alpha\beta\gamma}
+ \partial_i h_{\alpha\beta\gamma} - \left(
{\tilde K^{{ \delta} {\overline{\rho}} }}
\partial_i {\tilde K_{{\overline{\rho}}{ \alpha}}} h_{\delta\beta\gamma}
+(\alpha \leftrightarrow \beta)+(\alpha \leftrightarrow \gamma)\right)\right]
\end{eqnarray}
where
\begin{equation}
\label{fgrel1}
F^S = e^{G/2} K_{ {\bar{S}} S}^{-1} G_{\bar S}  \; \; , \;\;
F^i = e^{G/2} {\hat K}^{i {\overline j}} G_{\overline j}
\end{equation}
A generalization of the usual `angular parametrization' of
the F-field vev's will be introduced below in a representative example.
The matrix ${\hat K}^{i {\overline j}}$ is the inverse of
the moduli metric
${\hat K}_{ {\overline j} k}=\partial_{\overline j}\partial_k {\hat K}$, i.e.
${\hat K}^{i {\overline j}} {\hat K}_{ {\overline j} k} = \delta^i_k$.
Similarly, for the matter metric, we define
${\tilde K}^{\alpha {\overline \beta}}$ so that
${\tilde K}^{\alpha {\overline \beta}} {\tilde K}_{ {\overline \beta} \gamma}
= \delta^{\alpha}_{\gamma}$. Notice that, after normalizing the fields to get
canonical kinetic terms, the first piece in
eq.(\ref{mmatrix}) will lead to universal diagonal soft
masses but the second piece will generically induce
off-diagonal contributions. Concerning the
$A$-parameters, notice that in this section we have not
factored out the Yukawa couplings as usual, since
proportionality is not guaranteed.
Indeed, although the first term in
$A_{\alpha\beta\gamma}$ is always proportional
in flavour space to the corresponding Yukawa
coupling, the same thing is not necessarily true
for the terms contained in $\delta A$.
One purpose of this section is to study such `off-diagonal' effects
in the soft terms.

In order to get more concrete and manageable results, we will now
particularize the above formulae to
the untwisted sectors of $Z_3$,$Z_4$ and $Z_6'$ orbifolds.
The 9 $T^i$-moduli of the $Z_3$ orbifold enter in the K\"ahler potential as
elements of a $3\times 3$ matrix $T^{\alpha {\overline \beta}}$,
the role of the index $i$ being played by a pair of indices
(with $\alpha, {\overline \beta}=1,2,3$).
Similarly, the 4 $T^i$-moduli of $Z_4$ and $Z_6'$ orbifolds
associated to (say) the first and second complex planes
enter by a $2\times 2$ matrix $T^{\alpha {\overline \beta}}$
(with $\alpha, {\overline \beta}=1,2$). In addition, $Z_4$ ($Z_6'$)
has two additional moduli $T^3$ and $U^3$ (one additional modulus $T^3$)
associated to the third complex plane. Such moduli have diagonal metric,
as well as the associated untwisted fields. On the other side, the
moduli of `matrix' type and the associated untwisted charged fields
have non-diagonal metric, derivable from a K\"ahler potential of
the form
\begin{eqnarray}
\delta K & = & - \log \det\left( (T+T^{\dagger})^{\beta {\overline \alpha}}
- C^{\beta} {\overline C}^{\overline{\alpha}} \right)
\\
& \simeq  & - \log \det\ (T+T^{\dagger})^{\beta {\overline \alpha}}
+(T+T^{\dagger})^{-1}_{{\overline \alpha}\beta} {\overline
 C}^{\overline{\alpha}}
C^{\beta} \ .
\end{eqnarray}
It is convenient to define the hermitian matrix
\begin{equation}
t \equiv t^{\alpha {\overline \beta}} \equiv
(T+T^{\dagger})^{\alpha {\overline \beta}} \ .
\end{equation}
Then it is easy to find that the metric and inverse metric for moduli
and matter fields have the following simple expressions in terms of $t$:
\begin{equation}
\label{modmetr}
{\hat K}_{ {\overline i} j}=t^{-1}_{{\overline \alpha} \gamma }
t^{-1}_{{\overline \delta} \beta}  \;\; , \;\;
{\hat K}^{j {\overline i} }=t^{\gamma {\overline \alpha} }
t^{\beta {\overline \delta}}  \;\;
(i\equiv \alpha{\overline \beta} \; , \;  j\equiv\gamma{\overline \delta}) \ ,
\end{equation}
\begin{equation}
\label{matmetr}
{\tilde K}_{{\overline \alpha} \beta} = t^{-1}_{{\overline \alpha} \beta}
\;\; , \;\;
{\tilde K}^{\beta {\overline \alpha} } =t^{\beta {\overline \alpha} } \ .
\end{equation}
In addition, the $F^i$'s and $G_i$'s in such sectors are also conveniently
represented by matrices $F \equiv F^{\alpha {\overline \beta}}$ and
$G \equiv \partial{G}/ \partial T^{\alpha {\overline \beta}}$.
The relation between the matrices $F$ and $G$ follows from
eqs.~(\ref{fgrel1}) and (\ref{modmetr}):
\begin{equation}
\label{fgrel2}
F = m_{3/2} t G^* t \ .
\end{equation}

We first consider the $A_{\alpha\beta\gamma}$ parameters,
where the indices can now refer to any untwisted fields of
the orbifolds under study.
The relevant result is that $\delta A_{\alpha\beta\gamma}=0$.
This follows from the above structure of the metric and from the
antisymmetry property of Yukawa couplings with respect to
extra indices (understood above), e.g. $SU(3)$ indices in
(2,2) $Z_3$ orbifolds or $SU(2)$ indices in (2,2) $Z_4$, $Z'_6$
orbifolds. Therefore the result for $A_{\alpha\beta\gamma}$
is simply
\begin{equation}
A_{\alpha\beta\gamma} = F^S K_S h_{\alpha\beta\gamma}
= -\sqrt{3} m_{3/2} \sin\theta e^{-i{\gamma}_S} h_{\alpha\beta\gamma}
\label{ccc}
\end{equation}
which is the same result (after factorizing out the Yukawa coupling as
usual) as for the untwisted sector of
any other orbifold eq.(\ref{aaa}). Thus
{\it even in the presence of off-diagonal metrics
and multiple moduli} the result in eq.(\ref{aaa})
still holds.

We will now consider the soft mass matrix (\ref{mmatrix}) in one
of the sectors with off-diagonal metric. The result can be
written in the following compact form:
\begin{equation}
{\cal M}'^2 =
 m_{3/2}^2 t^{-1} - t^{-1} F t^{-1} F^{\dagger} t^{-1} \ .
\end{equation}
If the matter fields are canonically normalized as $C^{\alpha}
\rightarrow {\hat C}^{\alpha} = (t^{-1/2})^{\alpha}_{\beta} C^{\beta}$,
the normalized soft mass matrix can be written as
\begin{equation}
\label{mdelta}
{\cal M}^2 = m_{3/2}^2 ( 1 - \Delta) \ ,
\end{equation}
where $1$ stands for the unit matrix and the $\Delta$ is the matrix
\begin{equation}
\label{delmat}
\Delta = \frac{1}{m_{3/2}^2} t^{-1/2} F t^{-1} F^{\dagger} t^{-1/2} \ .
\end{equation}
It is interesting to notice that the contribution to SUSY-breaking
from the moduli of such a sector is
\begin{equation}
{\overline F}^{\overline i} {\hat K}_{ {\overline i} j } F^j
= m_{3/2}^2 {\mbox {Tr}} \Delta \ .
\end{equation}

To continue the discussion we will focus for definiteness
on the case of $Z_3$, where the 9 moduli $T^{\alpha {\overline \beta}}$
exhaust the set of untwisted moduli. We can consider the following
parametrization of the dilaton/moduli SUSY-breaking:
\begin{equation}
(S+S^*)^{-1} F^S = \sqrt{3} m_{3/2} \sin\theta e^{-i \gamma_S}
\;\; ; \;\;
t^{-1/2} F t^{-1/2}  = \sqrt{3} m_{3/2} \cos\theta \Theta \ ,
\end{equation}
where $\Theta$ is a $3\times 3 $ matrix satisfying
\begin{equation}
{\mbox  {Tr}} \Theta \Theta^{\dagger} = 1 \ .
\end{equation}
Notice that the matrix $\Delta$ in ${\cal M}^2$ (\ref{mdelta})
can be written
\begin{equation}
\label{deldel}
\Delta = 3 \cos^2\theta \, \Theta \Theta^{\dagger} \ .
\end{equation}
In particular, from this one immediately sees that: 1) $\Delta$ is
positive definite and ${\mbox {Tr}}\Delta=3 \cos^2\theta$  ;
2) the sum of the three eigenvalues of ${\cal M}^2$ satisfies
\begin{equation}
{\mbox {Tr}} {\cal M}^2 = 3 m_{3/2}^2 \sin^2\theta = |M|^2
\label{ddd}
\end{equation}
which confirms the already stated sum-rule eq.(\ref{rulix}) for untwisted
matter
in orbifolds, {\it even in the presence of off-diagonal metrics}.

An interesting question related to flavour changing issues\footnote{These 
were analyzed for the simplest case of diagonal metric in 
refs.\cite{BIM,LN}.}
concerns the degree of degeneracy among the three eigenvalues of ${\cal M}^2$.
It is clear that, for generic values (vev's) of the matrices $t$ and $F$
(or $\Theta$), $\Delta$ will have a generic matrix structure
and therefore the eigenvalues of ${\cal M}^2$ will be non-degenerate.
The approximately degenerate case occurs only when ${\cal M}^2$ is
approximately proportional to the unit matrix\footnote{This corresponds
to the simplest way of avoiding FCNC. Another possibility occurs
if scalar and fermionic mass matrices happen to be aligned \cite{NS}.
This and other issues on FCNC would require a detailed analysis of the flavour
structure of the models, which go beyond the scope of the present paper.},
i.e. ${\cal M}^2 \propto 1$.
This happens: 1) when $\Delta \ll 1$ ; 2) when $\Delta \propto 1$.

1) $\Delta \ll 1$. This happens when $\cos^2\theta \ll 1$, i.e.
when the contribution of the moduli $T^{\alpha {\overline \beta}}$
to SUSY-breaking is negligible. In the case of $Z_3$ this
just corresponds to the dilaton dominated SUSY-breaking
(in the case of $Z_4, Z_6'$ the SUSY-breaking could be shared between
$S$ and the third-complex-plane moduli). Actually, when discussing
FCNC constraints on soft masses, one should consider the
renormalization effects from the string scale to the electroweak
scale. Such effects include flavour independent contributions from
gauginos. For example, if squarks originated from a sector like
the one under study, the low energy mass matrix would read
${\cal M}^2(M_Z) \sim m_{3/2}^2 ( (1+ 24 \sin^2\theta) 1  - \Delta)$,
with $\Delta$ as in eq.(\ref{deldel}) for $Z_3$. 
Then the constraint $\cos^2\theta \ll 1$
would be relaxed to $\cos^2\theta \ll 1+ 24 \sin^2\theta$ \cite{BIM}
and the moduli would be allowed to participate to some extent to
SUSY-breaking. On the other side, no significant relaxation would be
obtained for sleptons.

2) $\Delta \propto 1$. This condition guarantees that ${\cal M}^2 \propto 1$
even when the moduli participate significantly to SUSY-breaking. Observing
eq.(\ref{delmat}), we can distinguish two subcases.
2a) If $t$ and $F$ are treated as independent objects, than the only
obvious way to satisy that condition is that both $t \propto 1$ and
$F \propto 1$. This requires not only that the off-diagonal moduli
and F-terms be negligible, but also that the diagonal ones be almost
identical, i.e. one is pushed towards the overall modulus limit.
2b) Such conclusion may be evaded if $t$ and $F$ are related
in some way, e.g. if $F \propto t$ (giving again $\Delta \propto 1$).
If this were the case, the off-diagonal elements of $F$ and $t$ would
not need to be negligible with respect to the diagonal ones.
An extreme example of this situation happens when $W$ does not depend
on the $T^{\alpha {\overline \beta}}$. In that case $F = -m_{3/2} t$
and $\Delta = 1$, implying ${\cal M}^2=0$ and a no-scale scenario.
An example where ${\cal M}^2 \neq 0$ can be obtained e.g. if $W$ depends
on $T^{\alpha {\overline \beta}}$ only via $\det T^{\alpha {\overline \beta}}$
(and if the vev of $T^{\alpha {\overline \beta}}$ is hermitian).

\section{The B-parameter and the $\mu $ problem}

When an (effective) $N=1$ SUSY mass
$\mu _{\alpha \beta }C^{\alpha }C^{\beta }$
appears in the Lagrangian of an $N=1$ theory,
SUSY-breaking also induces an associated
SUSY-breaking term
$B_{\alpha \beta }\mu_{\alpha \beta } C^{\alpha }C^{\beta }+h.c.$.
Very often these terms are absent due to gauge invariance. Thus in the
MSSM there is only one $B$-term associated to a
possible $\mu H_1H_2$ SUSY mass term. In fact both a $\mu$-term and a
$B$-term are phenomenologically required in the MSSM in order to, among other
things, avoid the presence of a visible axion.

The parameter $\mu $ of the MSSM has to be (on phenomenological grounds) of the
order of the low-energy SUSY-breaking scale (i.e., of order $m_{3/2}$).
The absence of a symmetry reason for such small value for $\mu $ is called
the ``$\mu $-problem" \cite{review}. Thus in order to be able to compute
$B$-term  in a given model, we need  first a mechanism which might naturally
induce a $\mu $-term of order $m_{3/2}$. We will discuss some of the
mechanisms proposed within the context of string-models to solve this
$\mu$-problem and we will also provide expressions for the associated
$B$-terms in this section.

\subsection{$B$-term from the K\"ahler potential in orbifold models}

It was pointed out in ref.\cite{GM} that terms in a K\"ahler potential
like the one proportional to $Z_{\alpha \beta }$ in eq.(\ref{kahl})
can naturally induce a $\mu $-term for the $C_{\alpha }$ fields
of order $m_{3/2}$ after SUSY-breaking, thus providing a rationale
for the size of $\mu $. Recently it has been realized that such type of
terms do appear in the K\"ahler potential of some Calabi-Yau type
compactifications \cite{KL}
and in orbifold models \cite{LLM,AGNT,FKZ}. Let us consider the case in which
e.g., due to gauge invariance, there is only one possible $\mu $-term
(and correspondingly one $B$-term) associated to a pair of matter fields
$C_1$,$C_2$. From eqs.(\ref{kahl},\ref{pot},\ref{auxi})
and from the fermionic part of the SUGRA lagrangian
one can check that a SUSY mass term $\mu C_1 C_2$
and a scalar term $B \mu (C_1 C_2) +h.c.$
are induced upon SUSY-breaking in the effective low energy theory
(here the kinetic terms for $C_{1,2}$ have been normalized to one).
If we introduce the abbreviations
\begin{equation}
L^Z \equiv  \log Z  \;\; , \;\;
L^{\alpha}  \equiv  \log {\tilde K}_{\alpha }  \;\; , \;\;
X  \equiv  1 - \sqrt{3}  \cos\theta \  e^{i\gamma _i}{\Theta _i}
({\hat K}_{ {\overline i} i})^{-1/2} L_{\overline i}^Z
\label{xxx}
\end{equation}
the $\mu$ and $B$ parameters (we will call them $\mu_Z$ and $B_Z$) are given by
\begin{eqnarray}
& \mu_Z \ =\ m_{3/2}  ( {\tilde K}_1 {\tilde K}_2 )^{-1/2} Z X \ , &
\label{mmu}
\\
& B_Z\ =\ m_{3/2} X^{-1}
\left[  2 + \sqrt{3} \cos\theta  ({\hat K}_{ {\overline i} i})^{-1/2}
{\Theta_i }
\left( e^{-i\gamma _i}
(  L_i^Z - L^1_i - L^2_i )
-e^{i\gamma _i} L_{\overline i}^Z  \right) \  \right.
& \nonumber\\
& \left. +
\ 3 \cos^2\theta ({\hat K}_{ {\overline i} i})^{-1/2}
{\Theta_i } e^{i\gamma _i} \ \left(
L_{\overline i}^Z ( L^1_j+L^2_j)
- L_{\overline i}^Z L_j^Z - L_{{\overline i} j}^Z\ \right)
({\hat K}_{ {\overline j} j})^{-1/2}
{\Theta _j } e^{-i\gamma _j}      \right] \ . &
\label{bcy}
\end{eqnarray}
%
The above formulae apply to the cases where the moduli on
which ${\tilde K}_1(T_i,T_i^*)$, ${\tilde K}_2(T_i,T_i^*)$ and
$Z(T_i,T_i^*)$ depend
have diagonal metric, which is the relevant case we are
going to discuss (anyway, the above formulae are easily generalized
to more general situations).

If the value of $V_0$ is not assumed to be zero, one just has to replace
$\cos\theta \rightarrow C\cos\theta$ in eqs.(\ref{xxx},\ref{mmu},\ref{bcy}),
where $C$ is given below eq.(\ref{gaugin}).
In addition, the formula for $B$ gets an additional contribution
given by $m_{3/2} X^{-1} 3(C^2-1)$.

It has been recently shown that the untwisted sector of orbifolds
with at least one complex-structure field  $U$  possesses the required
structure $Z(T_i,T_i^*)C_1C_2+h.c.$ in their K\"ahler
potentials. Specifically, the $Z_N$ orbifolds
based on $Z_4,Z_6$,$Z_8,Z_{12}'$ and the $Z_N\times Z_M$ orbifolds based
on $Z_2\times Z_4$ and $Z_2\times Z_6$ do all have a $U$-type field in (say)
the third complex plane. In addition the $Z_2\times Z_2$ orbifold has $U$
fields in the three complex planes.
In all these models the piece of the K\"ahler potential involving
the moduli and the untwisted matter fields $C_{1,2}$ in the third complex
plane has the form
\begin{eqnarray}
& K(T_i,T_i^*,C_1,C_2)=K'(T_l,T_l^*)
& \nonumber\\
& -\log\left((T_3+T_3^*)(U_3+U_3^*) - (C_1+C_2^*)(C_1^*+C_2)\right)
& \label{kahlb} \\
& \simeq
 K'(T_l,T_l^*)
 - \log(T_3+T_3^*)  - \log(U_3+U_3^*)\ +
\frac{(C_1+C_2^*)(C_1^*+C_2)}{(T_3+T_3^*)(U_3+U_3^*)}
\label{kahlexp}
\end{eqnarray}
The first term $K'(T_l,T_l^*)$ determines the (not necessarily diagonal)
metric of the moduli $T_l \neq T_3, U_3$ associated to the first and
second complex planes. The last term describes an
$SO(2,n)/SO(2)\times SO(n)$ K\"ahler manifold ($n=4$ if we
focus on just one component of $C_1$ and $C_2$) parametrized by
$T_3, U_3, C_1, C_2$. If the expansion shown in (\ref{kahlexp}) is
performed, on one hand one recovers the well known
factorization $SO(2,2)/SO(2)\times SO(2) \simeq (SU(1,1)/U(1))^2$
for the submanifold spanned by $T_3$ and $U_3$ (which have
therefore diagonal metric to lowest order in the matter fields),
whereas on the other hand one can easily identify the
functions $Z, {\tilde K}_1, {\tilde K}_2$ associated to $C_1$ and $C_2$:
\begin{equation}
Z\ =\ {\tilde K}_1 \ =\ {\tilde K}_2\ =\  {1\over {(T_3+T_3^*)(U_3+U_3^*)}}
\ .
\label{zzz}
\end{equation}
Plugging back these expressions in eqs.(\ref{mmu},\ref{bcy},\ref{xxx})
one can easily compute $\mu$ and $B$ for this interesting class
of models:
\begin{eqnarray}
& \mu_Z \ =\ m_{3/2}\ \left( 1\ +\ \sqrt{3}\cos\theta
(e^{i \gamma_3} \Theta _3 + e^{i \gamma_6} \Theta _6)\right) \ , &
\label{muu}
\\
& B_Z\mu_Z=2m_{3/2}^2 \left( 1+\sqrt{3} \cos\theta
 ( \cos\gamma_3 \Theta_3 + \cos\gamma_6 \Theta_6)  \  \right.
& \nonumber\\
& \left.
+\ 3\cos^2\theta \cos(\gamma_3-\gamma_6) {\Theta _3}{\Theta _6} \right) \ . &
\label{bmu}
\end{eqnarray}
In addition, we recall from eq.(\ref{untw}) that the soft masses are
\begin{equation}
m^2_{C_1}\ =\ m^2_{C_2}\ =\  m_{3/2}^2\ \left( 1\ -\ 3\cos^2\theta
(\Theta_3^2+\Theta _6^2)\right) \ .
\label{mundos}
\end{equation}
In general, the dimension-two scalar potential for $C_{1,2}$
(now denoting again normalized fields) after SUSY-breaking has
the form
\begin{equation}
 V_2(C_1,C_2)\ =\ (m_{C_1}^2+|\mu|^2)|C_1|^2 + (m_{C_2}^2+|\mu| ^2)|C_2|^2
+(B\mu C_1C_2+h.c.)\
\label{flaty}
\end{equation}
In the specific case under consideration, from
eqs.(\ref{muu},\ref{bmu},\ref{mundos}) we find the remarkable result,
which is also true for any value of $C$,
that the three coefficients
in $V_2(C_1,C_2)$ are equal, i.e.
\begin{equation}
m_{C_1}^2+|\mu_Z |^2 = m_{C_2}^2+|\mu_Z| ^2 = B_Z\mu_Z
\label{result}
\end{equation}
so that $V_2(C_1,C_2)$ has the simple form
\begin{equation}
V_2(C_1,C_2)\ =\  B_Z\mu_Z \ (C_1+C_2^*)(C_1^*+C_2) \ .
\label{potflat}
\end{equation}
Although the common value of the three coefficients in eq.(\ref{result})
depends on the Goldstino direction via the parameters
$\cos\theta$, $\Theta_3$, $\Theta_6$,\ldots (see expression of $B_Z\mu_Z$
in eq.(\ref{bmu})), we stress that the equality itself and the form
of $V_2$ hold {\em independently of the Goldstino direction}.
The only constraint that one may want to impose is that the coefficient
$B_Z\mu_Z$ be non-negative, which would select a region of parameter space.
For instance, if one neglects phases, such
requirement can be written simply as
\begin{equation}
(1+\sqrt{3} \cos\theta \ \Theta_3) (1+\sqrt{3} \cos\theta \ \Theta_6) \geq 0
\ .
\end{equation}
We notice in passing that the fields $C_{1,2}$ appear in the
SUSY-breaking scalar potential in the same combination as in
the K\"ahler potential. This particular form may be understood as due to
a symmetry under which $C_{1,2}\rightarrow C_{1,2}+i\delta $ in the K\"ahler
potential which is transmitted to the final form of the scalar potential.

An important (Goldstino-direction-independent) consequence of the above
form (\ref{potflat}) is that $V_2(C_1,C_2)$ identically vanishes
along the direction $C_1 = - C_2^*$, on which gauge symmetry is broken.
If dimension-four couplings respect such flat direction (which is
certainly the case for D-terms), we arrive at the important result
that along $<C_1>=-<C_2^*>$ the  {\it flatness is not
spoiled by the dilaton/moduli induced SUSY-breaking}.
This is certainly a very remarkable property.

This result can be rephrased in terms of the usual parameter
$\tan\beta=<C_2>/<C_1>$ (we now assume real vev's).
It is well known that, for a potential of the generic form
(\ref{flaty}) (+D-terms), the minimization conditions yield
\begin{equation}
\sin2\beta  \ =\ { {-2 B\mu} \over {m_{C_1}^2+m_{C_2}^2+2|\mu|^2} } \ .
\label{sbet}
\end{equation}
In particular, this relation embodies the boundedness requirement:
if the absolute value of the right-hand side becomes bigger than one,
this would indicate that the potential becomes unbounded from below.
As we have seen, in the class of models under consideration
the particular expressions of the mass parameters lead to
the equality (\ref{result}), which in turns implies
$\sin 2\beta= -1$. Thus one finds $\tan\beta =-1$
{\it for any value of $\cos\theta $,$\Theta _3 $,$\Theta _6 $} (and of
the other $\Theta_i$'s of course), i.e. for any Goldstino direction.

It is interesting to relate these results to similar ones
obtained in ref.\cite{BZ} in a slightly different context. In
ref.\cite{BZ} a {\em specific} SUGRA model was built,
where the Higgs-dependent part of the K\"ahler potential
had the form in eq.(\ref{kahlb}), with
$T_3=U_3$. The geometrical properties of the associated
manifold and a simple choice for the superpotential allowed
to obtain the simultaneous breaking of SUSY
and gauge symmetry, with the cosmological constant identically
vanishing along some flat directions which included the
$|C_1| = |C_2|$ direction. This also implied a partial participation
of charged fields in the process of SUSY-breaking\footnote{An elaboration
of this idea was later studied in ref.\cite{BFZ}.}. In the
limit of suppressed goldstino components along the Higgsinos,
SUSY-breaking was essentially dilaton/moduli dominated.
Then such model could be viewed as a very special
case of the more general framework here discussed,
characterized by specific values of the goldstino angles:
$\cos^2\theta=2/3$, $\Theta_3^2=\Theta_6^2=1/2$ and vanishing
values for the remaining $\Theta_i$'s. In particular
one had $V_2(C_1,C_2)\equiv 0$, the flat direction
$|C_1|=|C_2|$ being enforced by the D-term.
The remarkable result obtained in this section is that
the prediction $|\tan\beta|=1$ is actually valid for
a much broader class of models and holds irrespectively
of the goldstino direction in the dilaton/moduli space.
Whether the above mechanism can be successfully implemented
in the case of the electroweak Higgs fields remains
an open question. Flat potentials of the type here considered
could be interesting also for the breaking of a grand-unified
gauge group (as suggested e.g. in ref.\cite{BFZ}), in particular in
the context of models like string-GUTs \cite{AFIU}, in which
a vev of order the string scale is not problematic.

As an additional comment, it is worth recalling that in previous
analyses of the above mechanism for generating $\mu$ and $B$
in the string context \cite{KL,BLM,BIM} the value of $\mu$ was left
as a free parameter since one did not have an explicit expression for
the function $Z$. However, if the explicit orbifold formulae for
$Z$ are used, one is able to predict both $\mu$ and $B$ reaching
the above conclusion. We should add that situations are conceivable
where the above result may be evaded, for example if the physical Higgs
doublets are a mixture of the above fields with some other doublets coming
from other sectors (e.g. twisted) of the theory.

\subsection{$B$-term from the superpotential}



There is an alternative mechanism to the one studied in the previous subsection
to generate a $B$-term in the scalar potential.
It is well known that if the superpotential $W$ is assumed to have a
$\mu C_1C_2$ SUSY mass term, $\mu$ being an initial parameter, then a
$B$-term is automatically generated. We will call it $B_{\mu}$.
If we introduce the abbreviation
\begin{equation}
L^{\mu}  \equiv  \log {\mu}
\label{xxxx}
\end{equation}
the $\mu$ and $B$ parameters are given by
\begin{eqnarray}
& {\mu'} \ =\ {\mu} e^{K/2} \frac{W^*}{|W|}
({\tilde K}_1 {\tilde K}_2 )^{-1/2} \ , &
\label{mmuu}
\\
& B_{\mu}\ =\ m_{3/2}
\left[  -1 - \sqrt{3} e^{-i\gamma_S} \sin\theta (1- L_S^{\mu} 2ReS)
\right.
& \nonumber\\
& \left. +
\sqrt{3} \cos\theta  ({\hat K}_{ {\overline i} i})^{-1/2}
{\Theta_i } e^{-i\gamma _i}
({\hat K}_i + L_i^{\mu} - L^1_i - L^2_i ) \right] \ , &
\label{bcyy}
\end{eqnarray}
where the low-energy SUSY mass ${\mu'}$ is related to ${\mu}$ via
the usual SUGRA rescaling,
and again the kinetic terms for $C_{1,2}$ have been normalized to one. In
the above formulae we have assumed that in general ${\mu}$ will depend on
the SUSY-breaking sector fields, i.e. ${\mu}={\mu}(S,T_i)$.
These formulae are completely general and valid for any solution to the
${\mu}$-problem which introduces a small mass term $\mu(S,T_i) C_1C_2$ in $W$.
This type of solutions exists.

In ref.\cite{CM} was pointed out that the presence of a non-renormalizable
term in the superpotential
\begin{equation}
\lambda W C_1 C_2
\label{norenor}
\end{equation}
characterized by the coupling $\lambda$, yields dynamically a ${\mu}$ parameter
when $W$ acquires a vev
\begin{equation}
\mu = \lambda W \ .
\label{vev}
\end{equation}
The fact that $\mu$ is small is a consequence of our assumption of a correct
SUSY-breaking scale $m_{3/2}=e^{G/2}=e^{K/2}|W|$. The superpotential
eq.(\ref{norenor}) which provides a possible solution to the $\mu$ problem
can naturally be obtained in the context of strings.
A realistic example where non-perturbative SUSY-breaking mechanisms like
gaugino-squark condensation induce that superpotential was given in
ref.\cite{CM},
where $\lambda=\lambda(T_i)$ is a non-renormalizable Yukawa coupling between
the Higgses and the squarks and after eliminating the gaugino and squarks
bound states $W=W(S,T_i)$.
In ref.\cite{AGNT} the same kind of superpotential was obtained through
pure gaugino condensation in orbifolds with at least one complex-structure
field $U$. This is because in these orbifolds
matter field-dependent threshold corrections
($\propto C_1C_2$) appear in the gauge kinetic function $f$. We recall
that after eliminating the gaugino bound states the non-perturbative
superpotential $W\sim exp(3f/2b_0)$, where $b_0$ is the one-loop
$\beta$-function coefficient of the ``hidden'' gauge group.
After expanding the exponential, the superpotential will have a
contribution of the type (\ref{norenor}).
Again, $\lambda=\lambda(T_i)$, since the above
proportionality factor due to threshold corrections
depends on Dedekind functions which depend in turn on the moduli.

So with this solution (\ref{vev}) to the $\mu$-problem in strings:
\begin{equation}
\mu(S,T_i) = \lambda(T_i) W(S,T_i) \ .
\label{strings}
\end{equation}
Plugging back this expression in eqs.(\ref{mmuu},\ref{bcyy}) and imposing
the vanishing of the cosmological constant $V_0$, one can
easily compute $\mu$ and $B$ for this mechanism. We will call them
$\mu_{\lambda}$ and $B_{\lambda}$
\begin{eqnarray}
& \mu_{\lambda} \ =\ {\lambda} m_{3/2}
({\tilde K}_1 {\tilde K}_2 )^{-1/2} \ , &
\label{mmuuu}
\\
& B_{\lambda}\ =\ m_{3/2}
\left[  2 + \sqrt{3} \cos\theta  ({\hat K}_{ {\overline i} i})^{-1/2}
{\Theta_i } e^{-i\gamma _i}
(L_i^{\lambda} - L^1_i - L^2_i ) \right] \ , &
\label{blambda}
\end{eqnarray}
where
\begin{equation}
L^{\lambda}  \equiv  \log {\lambda} \ .
\label{xxxxx}
\end{equation}

If the value of $V_0$ is not assumed to be zero, one just has to replace
$\cos\theta \rightarrow C\cos\theta$ and
$\sin\theta \rightarrow C\sin\theta$
in eqs.(\ref{bcyy},\ref{blambda}),
where $C$ is given below eq.(\ref{gaugin}).
In addition, the formula for $B_{\lambda}$, eq.(\ref{blambda}),
gets an additional contribution
given by $m_{3/2} 3(C^2-1)$.

Concentrating again on the interesting case of orbifolds, where the
K\"ahler potential eq.(\ref{orbi}) is known, we obtain from eq.(\ref{blambda})
\begin{eqnarray}
& B_{\lambda}\ =\ m_{3/2}
\left[  2 - \sqrt{3} \cos\theta  \sum _{i=1}^6 e^{-i\gamma _i} {\Theta }_i
\left(n_1^i + n_2^i - \frac{\lambda_i}{\lambda} 2 Re T_i\right) \right] \ . &
\label{bcyyy}
\end{eqnarray}

Notice that it is conceivable that both mechanisms, the one solving the
$\mu$-problem through the K\"ahler potential (see subsection 4.1) \cite{GM}
and the other one solving it through the superpotential \cite{CM} shown above,
could be present simultaneously. In that case the general expressions for
$B$ and $\mu$ are easily obtained
\begin{eqnarray}
& \mu \ =\ \mu_Z + \mu_{\lambda} \ , &
\label{mmuuuuu}
\\
& B\ =\ \mu^{-1} (B_Z \mu_Z + B_{\lambda} \mu_{\lambda}) \ , &
\label{bcyyyy}
\end{eqnarray}
where $\mu_Z$, $B_Z$ are given in eqs.(\ref{muu},\ref{bmu}).
For example, in the case of orbifolds with at least one complex-structure
field $U$, where the $B_Z$-term from the K\"ahler potential is present, if
a gaugino condensate is formed, then automatically the $B_{\lambda}$-term
from the superpotential is also present as mentioned above. Now, as in the case
of $B_Z$ (see eqs.(\ref{muu},\ref{bmu})),
in $B_{\lambda}$ (\ref{bcyyy}) only ${\Theta}_3$ and ${\Theta }_6$
contribute. We recall that
the values of ${\tilde K}_1, {\tilde K}_2$ are given by eq.(\ref{zzz})
and besides, $\lambda=\lambda(T_3,U_3)$ (the concrete expression can be found
in ref.\cite{AGNT}). However, in this case the last equality of
eq.(\ref{result}) with $Z \rightarrow \lambda$ does not hold.

\section{Final comments and conclusions}

In this paper we have generalized in several directions
previous analyses of SUSY-breaking soft terms
induced by dilaton/moduli sectors. In particular, we
have studied the new features appearing when one goes to the
Abelian orbifold multimoduli case. We have found that there are qualitative
changes in the general patterns of soft terms. In some way
({\it on average}) the results are similar to the case in which
only $S$ and the ``overall modulus'' $T$ field are considered.
However, if one examines the soft terms for each particle individually
one finds different extreme patterns.
For example, non-universal soft scalar masses for particles with the
same overall modular weight are allowed and in fact this will be the
most general situation.
Besides, unlike in the
case considered in \cite{BIM}, gauginos may be lighter than
scalars even at the tree-level.
The possibilities are, however, not arbitrary. The fact that
{\it on average}
the results are similar to the simple $S,T$ case are embodied in
general sum rules like those in
eqs.(\ref{rulix},\ref{rulox},\ref{rulax},\ref{rulaxxt})
which relate soft terms of different particles in the theory.

Due to the mentioned sum-rules,
if we insist in obtaining results qualitatively different
from those in ref.\cite{BIM} (e.g., gauginos lighter than scalars
at the tree-level),
some scalars may get negative
mass$^2$.
This tachyonic behaviour may be just
signaling gauge symmetry breaking,
which might be a useful possibility in GUT model-building.
On the contrary, in the case of standard model 4-D strings,
the appearence of this tachyonic
behaviour could be dangerous since it could lead
to the breaking of charge and/or colour.
In order to avoid this problem, one is typically lead to a situation
with gauginos heavier than scalars, as in the overall modulus
case \cite{BIM}. We have also commented
on possible exceptions to such scenario (involving non renormalizable 
Yukawa couplings or negative soft mass$^2$ for the standard
model Higgses) which could lead to scalars heavier than gauginos.
Such inversion however can take place only for special goldstino
directions, and requires necessarily a small $\sin\theta$. We recall 
that the $\sin\theta \rightarrow 0$ limit was also the only one 
which could produce scalars heavier than gauginos in the overall 
modulus analysis, for other reasons (i.e. the different effect of 
string loop corrections on gaugino and scalar masses, vanishing
at tree-level).

We have also generalized our study to include the case of orbifolds with
off-diagonal untwisted $T^{\alpha {\overline \beta}}$
moduli. In this type of models
non-diagonal metrics for the untwisted matter fields appear. In spite
of this complication, sum rules analogous to those in
eqs.(\ref{rulix},\ref{aaa}) still hold (i.e., eqs.(\ref{ddd},\ref{ccc})).
Non-diagonal metrics for the matter fields do also in general
induce off-diagonal soft-masses for the scalars which in turn can induce
flavour-changing neutral currents depending on the size of the
off-diagonal moduli, as discussed in section 3.

We have finally considered the $\mu$ and $B$ terms obtained
in orbifold schemes. We have shown that the scheme in ref.\cite{GM}
in which a $\mu $-term is generated from a bilinear piece in the
K\"ahler potential, is rather constrained in its orbifold
implementation. We find that {\it irrespective of the Goldstino
direction} one always gets $|tg\beta |=1$ at the string scale.
Another way of stating the same result is that the flat direction
$\langle H_1 \rangle =\langle H_2 \rangle $ still remains flat
after including arbitrary dilaton/moduli-induced SUSY-breaking
terms. This is an intriguing result which could have interesting
 phenomenological applications. The results obtained for the
$B$-parameter in the scheme of ref.\cite{CM} in which a $\mu$-term
is generated from the superpotential are more model dependent.

A few comments before closing up are in order. First of all we are
assuming here that the seed of SUSY-breaking propagates through the
auxiliary fields of the dilaton $S$ and the moduli $T_i$ fields.
However attractive  this possibility might be, it is fair to say that
there is no compelling reason why indeed no other fields in the
theory could participate. Nevertheless the present scheme
has a certain predictivity due to the relative universality
of the couplings of the dilaton and moduli. Indeed, the
dilaton has universal and model-independent couplings which are
there independently of the four-dimensional string considered.
The moduli $T_i$ fields are less universal, their number and structure
depend on the type of compactification considered. However, there are
thousands of different $(0,2)$ models with different particle content
which share the same $T_i$ moduli structure. For example, the moduli
structure of a given $Z_N$ orbifold is the same for all the thousands
of $(0,2)$ models one can construct from it by doing different
embeddings and adding discrete Wilson lines. So, in this sense,
although not really universal, there are large classes of models with
identical $T_i$ couplings.
This is not the case of generic charged matter fields
whose number and couplings are completely out of control,
each individual model being in general completely different from any other.
Thus assuming dilaton/moduli dominance in the SUSY-breaking
process has at least the advantage of leading to specific
predictions for large classes of models whereas if charged
matter fields play an important role in SUSY-breaking we
will be forced to a model by model analysis,
something which looks out of reach.

Another point to remark is that we are using the tree level forms
for both the K\"ahler potential and the gauge kinetic function.
One-loop corrections to these functions have been computed in
some classes of four-dimensional strings and could be
included in the above analysis without difficulty.
The effect of these one-loop corrections will in general
be negligible except for those corners of the Goldstino
directions in which the tree-level soft terms vanish.
However, as already mentioned above, this situation would be a sort
of fine-tuning.
More worrysome are the possible non-perturbative
string corrections to the K\"ahler and gauge kinetic functions.
We have made use in our orbifold models of the
known tree-level results for those functions.
If the non-perturbative string  corrections turn out to be important,
it would be impossible to make any prediction about
soft terms unless we know all the relevant non-perturbative
string dynamics, something which looks rather remote
(although perhaps not so remote as it looked one year ago!).

One might hope that the relationships obtained among
soft terms in the dilaton/moduli dominated schemes
could be more general than the original tree-level
Lagrangians from which they are derived.
In this connection it has been recently realized that
the  boundary conditions $-A=M_{1/2}={\sqrt{3}}m$
of dilaton dominance coincide with some boundary
conditions considered by Jones, Mezincescu and Yau
in 1984 \cite{JMY}. They found that those same boundary conditions
mantain the (two-loop) finiteness properties of
certain $N=1$ SUSY theories. It has
 also been noticed \cite{I} that this
coincidence could be related to an underlying
$N=4$ structure of the dilaton Lagrangian
and that the dilaton-dominated boundary conditions
could also appear as a fixed point of renormalization group	
equations \cite{I,J}.
This could perhaps be an indication that at least
some of the possible soft terms obtained in
the present scheme could have a more general
relevance, not necessarily linked to
a particular form of a tree level Lagrangian.

\end{document}